\documentclass[10pt]{article}
\newcommand{\beq}{\begin{eqnarray}}
\newcommand{\eeq}{\end{eqnarray}}

\newcommand{\dev}{\mathrm d}
\newcommand{\I}{\imath}
\newcommand{\E}{\mathrm{e}^}

\newcommand{\im}{\mathrm{Im}}

\newcommand{\mP}{\mathcal{P}}
\newcommand{\mF}{\mathcal{F}}
\newcommand{\mG}{\mathcal{G}}

\usepackage{pstcol,times,color,pifont,graphicx,graphics,pstricks,pst-node,pst-coil}
\usepackage[latin1]{inputenc}

\def\vf_x{\stackrel{.}{f}_x}
\def\af_x{\stackrel{..}{f}_x}
\def\beq{\begin{equation}}
\def\eeq{\end{equation}}
\def\beqa{\begin{eqnarray}}
\def\eeqa{\end{eqnarray}}
\begin{document}
\title{Dispersion forces between an atom \\
and a perfectly conducting wedge}
\author{T N C Mendes$^{(a)}$
%\footnote{tarciro@if.uff.br}
 ,
 F S S Rosa $^{(b)}$
 %\footnote{darosa@lanl.gov}
  , A Ten\'orio$^{(c)}$
  %\footnote{tenorio@if.ufrj.br}
   and C Farina$^{(d)}$
   %\footnote{farina@if.ufrj.br}
\\ \textit{(a)  Universidade Federal Fluminense - Rio de Janeiro, RJ, 24 210-340, Brasil}\\
 \textit{(b)  Los Alamos National Laboratory, Los Alamos, NM 87545, USA}\\
  \textit{(c)  Centro Brasileiro de Pesquisas Físicas - Rio de Janeiro, RJ, 22 290-180, Brasil}
  \\
\textit{(d) Universidade Federal do Rio de Janeiro - Rio de Janeiro,
RJ, 21 941-972, Brasil}}
\date{}
 % Enter your date or \today between curly braces
\maketitle

\begin{abstract}

We consider the interaction between an electrically polarizable atom
in its fundamental state and a wedge constituted by two
semi-infinite perfectly conducting plates. Using a formalism based
on a master equation, we compute the dispersion force on the atom
for both retarded and non-retarded regimes.

\end{abstract}
%
%%%%%%%%%%%%%%%%%%%%%%%%%%%%%%%%%%%%%%%%%%%%%%%%%
%
\section{Introduction}

 Dispersion forces are generally understood as those
that occur  between neutral objects that do not possess any
permanent electric or magnetic moments, as the van der Waals force
between two neutral but polarizable atoms or between a polarizable
atom and a wall. These forces originates from the unavoidable
quantum fluctuations, which are always present in nature and can not
be neglected under certain circumstances. The dominant contribution
to the interaction of two neutral but polarizable objects comes from
the dipole-dipole interaction, the only one we shall be concerned in
this paper. Two distinct distance regimes are worth studying,
namely, the non-retarded regime (short distance regime) and the
retarded one (large distance regime). The occurrence of a dominant
transition wavelength naturally fixes a length scale which allows a
characterization for these two regimes. Retardation effects become
important as soon as typical distances between the two interacting
objects are of the order of the dominant transition wavelength.

Since the seminal papers by London \cite{London1930} (non-retarded
force between two polarizable atoms), by Casimir and Polder
\cite{CasimirPolder1948} (retardation effects on the London-van der
Waals force between two polarizable atoms and between an atom and a
perfectly conducting plate) and by Lifishitz \cite{Lifshitz1956}
 (who developed a general theory of dispersive van der Waals forces
between macroscopic dielectric bodies, valid also for $T\ne 0$)
 a wide knowledge about dispersion forces has been achieved: higher
multiple moments and N-atoms interactions have been considered;
non-additivity of these forces have been studied; the oscillatory
behaviour of the interaction between an excited atom and a wall has
been obtained \cite{Morawitz1969}; application of linear response
theory has been done \cite{McLachlan1963}, electric and magnetic
interactions have been treated in equal footing
\cite{Feinberg-Sucher970}; atom-cavity interaction has been
considered \cite{Barton1987,MeschedeEtAl1990}; thermal contribution
to the Casimir-Polder force has been computed
\cite{Goedecke-Wood1999}; the influence of the Casimir-Polder force
in Bose-Einstein condensates has been found \cite{LinEtAl2004,
AntezzaEtAl2004}; the importance of the Casimir-Polder force in
carbon nanotubes has been recognized \cite{BaglovEtAl2005};
for the computation of the force on a neutral atom near some
specific microstructures see C. Eberlein \cite{EberleinZietal2006}
(see also references therein); and many others.
 Nowadays, dispersion forces are important in many areas of science
 and finds applications in quite unexpected situations varying from
 biology, chemistry and physics to engineering and nanotechnology.
 For a recent  discussion on this subject from the point
of view of macroscopic QED in linear media, which contains a
detailed historical survey as well as a vast list of references, see
\cite{BuhmannWelschPQE2007} and references therein. The first
experiment conceived to measure directly the atom-body interaction
was done by Sukenik {\it et al} in 1993 \cite{SukenikEtAl1993}.
These authors analised the deflection of ground-state sodium atoms
crossing a micron-sized parallel-plate cavity (in fact, a wedge with
very small angle). The authors conclusively confirmed the existence
of retardation effects. There are other experiments as, for example,
the Orsay experiment, which measures the force between an atom and a
dielectric wall by analyzing the reflection of atoms by
evanescent-wave atomic mirrors \cite{LandraginEtAl1996}, and the
Tokyo experiment, based on quantum reflection by the Casimir-Polder
force \cite{Shimizu2001} and even measurements that employ the
influence of dispersion forces on BEC \cite{CornelExperiment}

Our purpose is to obtain the dispersion force, in any distance regime, exerted on an
electrically polarizable atom in ground state, which is near a wedge formed by two
perfectly conducting plates of infinite extent. The atom-wedge
system has been considered before by Brevik, Lygren and Marachevsky
\cite{BrevikEtAl1998} but these authors computed the dispersion
force on the atom only in the retarded regime. They based their
calculations in a previous work on the Casimir effect for a
perfectly conducting wedge \cite{BrevikEtAl1996} and the inclusion of a
dielectric was done in \cite{BrevikEtAl2001}. Since the wedge
geometry was employed in the first measurement of the Casimir-Polder
force \cite{SukenikEtAl1993} and it may be convenient in
future ones, we think it is of some relevance to generalize the results
in \cite{BrevikEtAl1998} by providing the calculation of the
dispersion force also in the non-retarded regime.
 \section{Dispersion potential for the atom-wedge system}

Consider an atom and a perfectly conducting wedge as indicated in
Figure 1, where $\rho$ is the distance between the atom and the
corner, chosen as the ${\cal OZ}$ axis, and $\phi$ is the polar
angle, measured with respect to the ${\cal OXZ}$ plane.
 We shall employ a method developed by Dalibard, Dupon-Roc and Cohen-Tannoudji
 \cite{DalibardEtAl1984}, based on a master equation to describe
 a small system interacting  with a large one (refered to as a reservoir).
 In our case the small system will be the atom, while the large one will be the
 radiation field submitted to the appropriate boundary conditions
 imposed by the wedge. This approach provides general expressions for the level
shifts and energy exchange rates of the system. Two contributions
appear: one from the fluctuation of the reservoir, denoted by $fr$,
and the other from the reaction of the reservoir, denoted by $rr$.

  The level shifts $\delta E_a$
  %
  %and energy exchange rates  $\dot{\mathcal Q}_a$
  %
  for an atom in the state $\vert a\rangle$ interacting with the radiation field in the
dipole approximation are given by \cite{TarciroJPA}
%

%%%%%%%%%%%%%%%%%%%     AQUI COMECA A UNICA FIGURA EM LATEX      %%%%%%%%%%%%%%%%%%%%%%%%

%%%%%%%%%%%%%%%%%%   Atomo - Cunha   %%%%%%%%%%%%%%%%%%%%%%
%

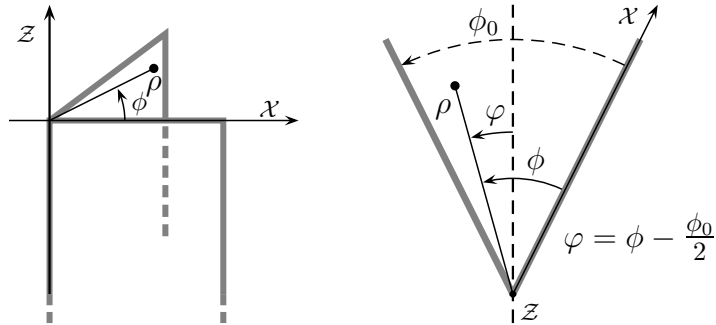
\begin{figure}[h!]
\begin{center}
\newpsobject{showgrid}{psgrid}{subgriddiv=1,griddots=10,gridlabels=6pt}
\psset{unit=0.77}
\begin{pspicture}(-9,-0.2)(3,4.9)
%\showgrid
 \psset{arrowsize=0.12 2}
%
%%%  Eixo Horizontal
%\psline[linewidth=.3mm]{<-}(-2.5,0)(2.5,0)
% \rput(-2.0,-0.4){\small${\cal Y}$}
 %%%  Letra Z
 \rput(0.3,-0.3){\small${\cal Z}$}
%

%%%%   Linha Vertical pontilhada
\psline[linewidth=.3mm,linestyle=dashed]{-}(0,-0.5)(0,5)
%
%%%   Placas formando a cunha
\psline[linewidth=0.9mm,linecolor=gray]{-}(0,0)(2.2,4.4)
\psline[linewidth=0.9mm,linecolor=gray]{-}(0,0)(-2.2,4.4)
 %%%  Articulacao na origem
 \pscircle[fillstyle=solid,fillcolor=black](0,0){.06}
 %%%   Angulo phi0
  \psarc[linewidth=.2mm,linestyle=dashed]{->}(0,0){4.4}{64}{116}
 \rput(-0.5,4.7){\large$\phi_0$}

%%%%   Eixo OX
\psline[linewidth=.2mm]{->}(0,0)(2.5,5)
 \rput(2,4.8){\small${\cal X}$}

 %%%   Atomo
 \pscircle[fillstyle=solid,fillcolor=black](-1,3.6){.08}
\psline[linewidth=.2mm,linecolor=black]{-}(0,0)(-1,3.6)
 %%%  Angulo phi
 \psarc[linewidth=.2mm,linecolor=black]{->}(0,0){2.0}{65}{105}
 \rput(0.4,2.3){\large$\phi$}

 %%%   Angulo varphi
 \psarc[linewidth=.2mm,fillcolor=black]{->}(0,0){2.8}{90}{105}
 \rput(-0.3,3.1){\large$\varphi$}
 \rput(-1.2,3.2){\large$\rho$}
 %
 %
%%%%%%%%%%%%%%%%%%%%%
%
%

%%%  Vista em perspectiva  Figura da esquerda
\psline[linewidth=0.8mm,linecolor=gray]{-}(-8,0)(-8,3)(-5,3)(-5,0)
\psline[linewidth=0.8mm,linecolor=gray,linestyle=dashed]{-}(-8,0)(-8,-0.5)
\psline[linewidth=0.8mm,linecolor=gray,linestyle=dashed]{-}(-5,0)(-5,-0.5)
\psline[linewidth=0.8mm,linecolor=gray]{-}(-8,3)(-6,4.5)(-6,3)
\psline[linewidth=0.8mm,linecolor=gray,linestyle=dashed]{-}(-6,2.9)(-6,1)
 %%%   Eixo Vertical, Z
 \psline[linewidth=.3mm]{->}(-8,0)(-8,5)
 \rput(-8.4,4.5){\small${\cal Z}$}
 %%%   Eixo X
 \psline[linewidth=.2mm]{->}(-8.7,3)(-3.7,3)
 \rput(-4.2,3.2){\small${\cal X}$}
 %%%   Atomo entre placas na esquerda
  \pscircle[fillstyle=solid,fillcolor=black](-6.2,3.9){.08}
 \psline[linewidth=.2mm,linecolor=black]{-}(-8,3)(-6.2,3.9)
 \rput(-6.2,3.6){\large$\rho$}
 %%%   Angulo phi
 \psarc[linewidth=.2mm,linecolor=black]{->}(-8,3){1.3}{0}{25}
 \rput(-6.45,3.29){$\phi$}

 %%%  Formula: relacao entre varphi e phi
  \rput(2.2,1){\large$\varphi = \phi - \mbox{\Large$\frac{\phi_0}{2}$}$}
 %%%   Eixo OY
 %\psline[linewidth=.3mm]{->}(-7.3,2.2)(-8.7,3.8)
%
%
\end{pspicture}
\caption{Atom-wedge system:  the figure on the right shows a
transverse cut.}
\end{center}
\label{Figure1}
\end{figure}
%
%

%%%%%%%%%%%%%%%     AQUI TERMINA A UNICA FIGURA EM LATEX    %%%%%%%%%%%%%%%%

%
%
${\;}$

${\;}$

\vskip -2cm

\begin{eqnarray}
\label{dQa}
%
%\dot{\mathcal Q}_a &=& \dot{\mathcal Q}^{rr}_a + \dot{\mathcal
%Q}^{fr}_a\,,
%%
%%\\
%%
%%%%%%%%%%%%%%%%%%%%%%%%%%%%%%%%%%%%%%%%%%%%%%%%%%%%%%%%%%%%%%%%%%%%%%%%%%%%
%%\label{dEa}
%%
%\;\; ;\;\;\;\;\;\;\;\;\;\;\;
 \delta E_a &=& \delta E^{rr}_a + \delta E^{fr}_a\,,
\\
%
%%%%%%%%%%%%%%%%%%%%%%%%%%%%%%%%%%%%%%%%%%%%%%%%%%%%%%%%%%%%%%%%%%%%%%%%%
%
\label{dEa_rr}
\delta E^{rr}_a &=& -{1\over 2}\sum_j
\sum_{\mathbf{k}\lambda}\alpha_{aj}^{\,\prime(-)}
 \left(k\right)\vert f^j_{\mathbf{k}\lambda}\left(\mathbf{x}\right)\vert^2 \,,
\\
%%%%%%%%%%%%%%%%%%%%%%%%%%%%%%%%%%%%%%%%%%%%%%%%%%%%%%%%%%%%%%%%%%%%%%%%
%%
%\label{Qa_rr}
%%
%\dot{\mathcal Q}_a^{rr} &=& - \sum_j
%\sum_{\mathbf{k}\lambda}c\,k\,\alpha_{aj}^{\,\prime\prime(-)}
% \left(k\right)\vert f^j_{\mathbf{k}\lambda}\left(\mathbf{x}\right)\vert^2 \,,
%%
%\\
%%%%%%%%%%%%%%%%%%%%%%%%%%%%%%%%%%%%%%%%%%%%%%%%%%%%%%%%%%%%%%%%%%%%%%%
%
\label{dEa_fr}
\delta E^{fr}_a &=& -\sum_j \sum_{\mathbf{k}\lambda}
\alpha_{aj}^{\,\prime(+)}
 \left(k\right)\vert f^j_{\mathbf{k}\lambda}\left(\mathbf{x}\right)\vert^2
 \left( \langle n_{\mathbf{k}\lambda}\rangle + {1\over 2}\right)\,,
\\
%%%%%%%%%%%%%%%%%%%%%%%%%%%%%%%%%%%%%%%%%%%%%%%%%%%%%%%%%%%%%%%%%%%%%%%%%%%%%%%%
%%
%\label{Qa_fr}
%%
%\dot{\mathcal Q}_a^{fr} &=& \sum_j \sum_{\mathbf{k}\lambda}
%c\,k\,\alpha_{aj}^{\,\prime\prime(+)}\left(k\right)\vert
%f^j_{\mathbf{k}\lambda}
% \left(\mathbf{x}\right)\vert^2 \Big( 2\langle n_{\mathbf{k}\lambda}\rangle + 1\Big)\,,
%%
%\\
%%%%%%%%%%%%%%%%%%%%%%%%%%%%%%%%%%%%%%%%%%%%%%%%%%%%%%%%%%%%%%%%%%%%%%%%%
%
\label{a'+-}
\alpha_{aj}^{\,\prime(\mp)}\left(k\right) &=& \sum_b{\alpha_{ab}^{j}
k_{ba}\over 2}
 \left[ \mP {1\over k + k_{ba}}\; \pm \; \mP {1\over k-k_{ba}}\right]\,,
%
%\\
%%%%%%%%%%%%%%%%%%%%%%%%%%%%%%%%%%%%%%%%%%%%%%%%%%%%%%%%%%%%%%%%%%%%%%%%%%%%%%%%%
%%
%\label{a''+-}
%%
%\alpha_{aj}^{\,\prime\prime(\mp)}\left(k\right) &=&
%\pi\sum_b{\alpha_{ab}^{j} k_{ba}\over 2}
% \Big[ \delta\left( k - k_{ba}\right)\pm\delta\left( k + k_{ba}\right)\Big]\,,
 %
\end{eqnarray}
with the electric field being
%
%is written in the form
%
$ \mathbf{E}\left(\mathbf{x},t\right) =
\sum_{\mathbf{k}\lambda}\left(\mathbf{f}_{\mathbf{k}\lambda}\left(\mathbf{x}\right)\E{\I\omega_k
t} a_{\mathbf{k}\lambda}^{\dag}+\mathbf{f}_{\mathbf{k}\lambda}^{*}
\left(\mathbf{x}\right)\E{-\I\omega_k t}
a_{\mathbf{k}\lambda}\right), $
 where
 $ \omega_k = k c, \;
         \big[a_{\mathbf{k}\lambda},a_{\mathbf{k}^{\prime}\lambda^{\prime}}\big]
=
 \big[a_{\mathbf{k}\lambda}^{\dag},a_{\mathbf{k}^{\prime}\lambda^{\prime}}^{\dag}\big] = 0\;,\;
%
%\\
%%%%%%%%%%%%%%%%%%%%%%%%%%%%%%%%%%%%%%%%%%%%%%%%%%%%%%%%%%%%%%%%%%
%\label{Com_aa+}
%
\big[a_{\mathbf{k}\lambda},a_{\mathbf{k}^{\prime}\lambda^{\prime}}^{\dag}\big]
=
 \delta_{\mathbf{k}\mathbf{k}^{\prime}}\delta_{\lambda\lambda^{\prime}}\;
 ,
$
 $\mP$ means the principal part, $k_{ab}c$ is the transition
frequency between states $\vert a\rangle$ and $\vert b\rangle$,
$\langle n_{{\bf k}\lambda}\rangle$ is the average number of photons
in the mode ${\bf k}\lambda$
%
%(which depends on the field state),
%
 and coefficients $\alpha_{ab}^j$ are defined as
  $
  \alpha_{ab}^j =
- {2\vert\langle a\vert d_j\vert b\rangle\vert^2}/(\hbar c \,
k_{ab})\, ,
 $
with $d_j$ being the $j$-component  of its dipole moment operator
$ \mathbf{d} = -e\mathbf{r} = -e\left(x_1,x_2,x_3\right) =
\left(d_1,d_2,d_3\right)\, .
 $
For convenience, we first consider the field modes of a wedge with a
coaxial cylindrical shell of radius $R$ which will be taken to
infinity at the appropriate moment. The transverse electric (TE) and
transverse magnetic (TM) modes of the quantized electric field for
this set up are borrowed from A. A. Saharian's paper \cite{Saharian2007}
and are given by
\begin{eqnarray}
\label{TM}
\mathbf{f}_{\mathbf{k},m,n}^{\mathrm{TM}}\!\!\left(\mathbf{x}\right)\!\!\!\!\!
&=&\!\!\!\!\!\! \beta_{qm}\!\!\left(\gamma_{\vert
m\vert,n}R\right)\!\left(\gamma_{\vert m\vert,n}^2\hat{z} - \imath
k_z\mathbf{\nabla}_t\right)\! J_{q\vert
m\vert}\!\!\left(\gamma_{\vert m\vert,n}\rho\right)\!\sin\left(q
\vert m\vert \phi\right)\! e^{-\imath\left(k_zz-\omega_k
t\right)}\;\;\;\;\;
\cr
%%%%%%%%%%%%%%%%%%%%%%%%%%%%%%%%%%%%%%%%%%%%%%%%%%%%%%%%%%%%%%%%%%%
%
\label{TE}
\mathbf{f}_{\mathbf{k},m,n}^{\mathrm{TE}}\left(\mathbf{x}\right)\!\!\!\!\!
&=&\!\!\!\!\! \imath k\beta_{qm}\left(\eta_{\vert
m\vert,n}R\right)\hat{z}\times\mathbf{\nabla}_t\left[J_{q\vert
m\vert}\left(\eta_{\vert m\vert,n}\rho\right)\cos\left(q \vert
m\vert \phi\right)e^{-\imath\left(k_zz-\omega_k
t\right)}\right]\nonumber
\end{eqnarray}
 where
$
 J_{q\vert m\vert}\!\left(\gamma_{\vert
m\vert,n}R\right) \!=\! J^{\,\prime}_{q\vert
m\vert}\!\left(\eta_{\vert m\vert,n}R\right) \!= 0,\;X_{\nu}(x)\! =
 \!\!\left[J^{\,\prime\,2}_{\nu}\left(x\right) +
 \!\left(1-\mbox{\large$\frac{\nu^2}{x^2}$}\right)
 \!J_{\nu}^2\left(x\right)\right]^{-1}\, ,
 $
 $
\;\; {\beta}_{qm}^2\left(x\right) =
 \mbox{\large$\frac{2q\hbar c}{\pi k}$}\,X_{qm}\left(x\right),
 $
$
 \mathbf{\nabla}_t = \hat\rho\,\partial_{\rho} +
 \mbox{\large$\frac{1}{\rho}$}\,\hat\phi\,\partial_{\phi},
 \;
 q = {\pi}/{\phi_0}, \; k^2 = \kappa_{mn\lambda}^2+k_z^2
  \, ,
  $
  $
  %\linebreak
   \kappa_{mn1} = \gamma_{\vert m\vert,n}\;
   $
   and
   $
 \;\kappa_{mn2} = \eta_{\vert m\vert,n}\, .
 $
The atom-boundary interaction is given by the position-dependent
part of the energy shift induced by the atom-field interaction (with
appropriate BC). Computing first the $(rr)$ contribution, we have
%\vskip -0.2cm
%
$$
%
%\label{dE_as}
%
\delta E_{a}^{rr} =
 \! -{q\hbar c\over\pi}\!\!\!\int_{-\infty}^{\infty}
 \!\!\!\!\!\dev k_z
 \!\!\!\!\!\!\sum_{m=-\infty}^{\infty}
 \!\sum_{n=1}^{\infty}\!\sum_{\lambda,\sigma}\!{\kappa_{mn\lambda}^4
 X_{qm}\left(\kappa_{mn\lambda}R\right)\over
\sqrt{\kappa_{mn\lambda}^2+k_z^2}}\alpha_{a\sigma}^{\,\prime\,(-)}
 \!\!\!\left(\!\!\sqrt{\kappa_{mn\lambda}^2+k_z^2}\right)
 \! Q_{qmn}^{\,\sigma,\lambda}\left(\rho,\phi\right)\;\;\;\;
$$
where $\sigma = \rho, \phi\, z$, the index
 $\lambda = 1,2$
reffers to TM and TE modes respectively and
\begin{eqnarray}
\label{Q_pTM}
Q_{qmn}^{\,\rho,1}\left(\rho,\phi\right)&=&
{k_z^2\over\kappa_{mn1}^2}J_{qm}^{\,\prime\,2}\left(\kappa_{mn1}\rho\right)\sin^2\left(qm\phi\right)\,,
\cr
\label{Q_fiTM}
Q_{qmn}^{\,\phi,1}\left(\rho,\phi\right)&=&{k_z^2q^2 m^2\over
\kappa_{mn1}^4\rho^2}
J_{qm}^2\left(\kappa_{mn1}\rho\right)\cos^2\left(qm\phi\right)\,,
\cr
\label{Q_zTM}
Q_{qmn}^{\,z,1}\left(\rho,\phi\right)&=&
J_{qm}^2\left(\kappa_{mn1}\rho\right)\sin^2\left(qm\phi\right)\,,
 \;\;\;\;Q_{qmn}^{\,z,2}\left(\rho,\phi\right)\;\;=\;\;0\,,
\cr
\label{Q_fiTE}
Q_{qmn}^{\,\phi,2}\left(\rho,\phi\right)&=&\left(1+{k_z^2\over\kappa_{mn2}^2}\right)
 J_{qm}^{\,\prime\,2}\left(\kappa_{mn2}\rho\right)\cos^2\left(qm\phi\right)\,,
\cr
\label{Q_pTE}
Q_{qmn}^{\,\rho,2}\left(\rho,\phi\right)&=&\left(1+{k_z^2\over\kappa_{mn2}^2}\right){q^2
m^2\over \kappa_{mn2}^2\rho^2}
J_{qm}^2\left(\kappa_{mn2}\rho\right)\sin^2\left(qm\phi\right)
 \, ,\nonumber
\end{eqnarray}
Using the generalized Abel-Plana summation formula
 and taking {$R\rightarrow\infty$}, we get \cite{Saharian2007}
$$
%
%\label{dErr}
%
\delta E_{a}^{rr}= -{q\hbar c\over
2\pi}\sum_{m=-\infty}^{\infty}\int_{-\infty}^{\infty}dk_z\int_{0}^{\infty}{\kappa^3
d\kappa\over\sqrt{k_z^2+\kappa^2}}\sum_{\sigma}\alpha_{a\sigma}^{\,\prime(-)}
 \left(\sqrt{k_z^2+\kappa^2}\right)S_{\kappa
qm}^{\,\sigma}\left(\rho,\phi\right)\, ,
%%%%%%%%%%%%%%%%%%%%%%%%%%%%%%%%%%%%
$$
\vskip -0.5cm
\begin{eqnarray}
\label{Sz}
S_{\kappa qm}^{\,z}\left(\rho,\phi\right)&=&
J_{qm}^2\left(\kappa\rho\right)\sin^2\left(qm\phi\right)\,,
\cr
%
%%%%%%%%%%%%%%%%%%%%%%%%%%%%%%%%%%%%%%%%%%%%%%%%%%%%%%%%%%%%%%%%%%%%%%%%%%%%%%%%%%%%%%%%
%
\label{Sfi}
S_{\kappa qm}^{\,\phi}\left(\rho,\phi\right)&=&
\left[\left(1+{k_z^2\over\kappa^2}\right)J_{qm}^{\,\prime\,2}
 \left(\kappa\rho\right)+{k_z^2q^2m^2\over\kappa^4\rho^2}J_{qm}^2
 \left(\kappa\rho\right)\right]\cos^2\left(qm\phi\right)\,,
\cr
%
%%%%%%%%%%%%%%%%%%%%%%%%%%%%%%%%%%%%%%%%%%%%%%%%%%%%%%%%%%%%%%%%%%%%%%%%%%%%%%%%%%%%
%
\label{Sp}
S_{\kappa qm}^{\,\rho}\left(\rho,\phi\right)&=&
\left[{k_z^2\over\kappa^2}J_{qm}^{\,\prime\,2}
 \left(\kappa\rho\right)+\left(1+{k_z^2\over\kappa^2}\right){q^2m^2\over\kappa^2\rho^2}
 J_{qm}^2\left(\kappa\rho\right)\right]\sin^2\left(qm\phi\right)\,\nonumber
%
%%%%%%%%%%%%%%%%%%%%%%%%%%%%%%%%%%%%
\end{eqnarray}
There is no need for a cut off function: the polarizabity guarantees
the convergence of the sums for the position-dependent part of
{$\delta E_a^{(rr)}$}. It can be shown by explicit calculations that
the {$(fr)$} contribution is obtained by
 $
 \alpha_{a\sigma}^{\,\prime
(-)}(k)\longrightarrow\alpha_{a\sigma}^{\,\prime
(+)}(k)\left(2\langle n_{{\bf k}\lambda}\rangle+1\right)\, . $
 A futher simplification is possible
only for positive integer values of  $q,\; (q=1,2,3...)$. Using a
generalization of the addition theorem for Bessel functions
\cite{DaviesSahni1988},
$$
\sum_{m=-\infty}^{\infty}J_{qm}\left(\kappa\rho\right)Z_{\nu+qm}\left(\kappa\rho\right)\E{2\I
qm\phi} = {1\over
q}\sum_{l=0}^{q-1}\left(-1\right)^{\nu/2}\E{-\I\nu\psi_l}Z_{\nu}\left(2\kappa\rho\sin\psi_l\right)\,,\nonumber
$$
where $Z_{\nu}$ is a solution of Bessel's equation and $\psi_l =
\phi+\vartheta_l, \;\vartheta_l = \pi l/q$, the
 %\linebreak
 sums over $m$ can be evaluated. Passing to
spherical coordinates ($\kappa = k\sin\theta$,
% \linebreak
 $k_z = k\cos\theta$) and making
 use of identities involving Bessel functions, we get
\begin{eqnarray}
\label{dEaz_rr}
\delta E_{a,z}^{rr} \!\!\!\!&=&\!\!\!\! {\hbar c\over
2\pi}\sum_{l=0}^{q-1}\int_{0}^{\infty} \!\!\! dk\, k^3
\alpha_{az}^{\,\prime(-)}\!(k)
\Big[G_{\parallel}\left(2k\rho\sin\psi_l\right) -
G_{\parallel}\left(2k\rho\sin\vartheta_l\right)\Big]\,,
\cr
%%%%%%%%%%%%%%%%%%%%%%%%%%%%%%%%%%%%%
%
\label{dEafi_rr}
\delta E_{a,\phi}^{rr} \!\!\!\!&=&\!\!\!\! {\hbar c\over
2\pi}\sum_{l=0}^{q-1}\int_{0}^{\infty} \!\!\! dk\, k^3
\alpha_{a\phi}^{\,\prime(-)}\!(k) \Big[
H_{\phi}\left(2k\rho,\psi_l\right)+H_{\phi}\left(2k\rho,\vartheta_l\right)\Big]\,,
\cr
%%%%%%%%%%%%%%%%%%%%%%%%%%%%%%%%%%%%%
%
\label{dEarho_rr}
\delta E_{a,\rho}^{rr} \!\!\!\!&=&\!\!\!\! {\hbar c\over
2\pi}\sum_{l=0}^{q-1}\int_{0}^{\infty}\!\!\! dk\, k^3
\alpha_{a\rho}^{\,\prime(-)}\!(k)\Big[
H_{\rho}\left(2k\rho,\psi_l\right)-H_{\rho}\left(2k\rho,\vartheta_l\right)\Big]\,,\nonumber
%
%\cr
%
%
%%%%%%%%%%%%%%%%%%%%%%%%%%%%%%%%%%%%
\end{eqnarray}
where we defined
\begin{eqnarray}
\label{Gper}
G_{\perp}(x) &=& {\cos x\over x^2}-{\sin x\over x^3}\; ;\;\;\;\;
%
%\cr
%%%%%%%%%%%%%%%%%%%%%%%%%%%%%%%%%%%%%%%%%%%%%%%%%%%%%%%%%%%%%%%%%%%%%%%%%%
%
\label{Gparl}
G_{\parallel}(x) = {\sin x\over x}+{\cos x\over x^2}-{\sin x\over
x^3}\,,
\cr
%%%%%%%%%%%%%%%%%%%%%%%%%%%%%%%%%%%%%%%%%%%%%%%%%%%%%%%%%%%%%%%%%%%%%%%%%%
%
\label{H1}
H_{\phi}\left(x,\psi\right)&=&G_{\parallel}\left(x\sin\psi\right)
 \sin^2\psi+2G_{\perp}\left(x\sin\psi\right)\cos^2\psi\,,
\cr
%%%%%%%%%%%%%%%%%%%%%%%%%%%%%%%%%%%%%%%%%%%%%%%%%%%%%%%%%%
%
\label{H2}
H_{\rho}\left(x,\psi\right)&=&G_{\parallel}\left(x\sin\psi\right)
 \cos^2\psi+2G_{\perp}\left(x\sin\psi\right)\sin^2\psi\, . \nonumber
%
%%%%%%%%%%%%%%%%%%%%%%%%%%%%%%%%%%%%%%%%%%%%%%%%%%%%%%%%%%%%%%%%%
%
\end{eqnarray}
We still need to evaluate the {$k$-integrals}. From the expressions
for the {$\alpha_{a\sigma}^{\,\prime}(k)$}, these integrals have the
form
%
%\begin{eqnarray}
%
%\label{A+-}
%
$
\mathcal{A}^{(\pm)}_{\lambda}\left(\omega, f\right) =
 \int_{0}^{\infty}\dev k\,f\left(k\right)
 \left[\mP \mbox{\large$\frac{1}{k + \omega}$} \mp\mP
  \mbox{\large$\frac{1}{k - \omega}$} \right]\E{\imath k\lambda}\, ,
 %\nonumber
%
$
%\end{eqnarray}
%
where $\omega>0$ and $\lambda > 0$ is a real parameter. For an
analytical function $f$ satisfying, in the whole complex plane, the
condition
 $\vert f\left(k\right)\vert \E{-\lambda\vert\im\left[k\right]\vert}
  \rightarrow  0$,
 for $\vert\im\left[k\right]\vert\rightarrow\infty$,
we get
$
\mathcal{A}^{(\pm)}_{\lambda}\left(\omega, f\right) =
 \mp\I\pi f\left(\omega\right)\E{\imath \omega\lambda} +
 \int_{0}^{\infty}{\dev k\over k+\omega}\left[f\left(k\right)\E{\imath k\lambda}\mp
 f\left(-k\right)\E{-\imath k\lambda}\right].
 %\;\;\;\;\;\;\;
%
$
Hence, the $(rr)$ contributions to the potential for each
polarizations are given by
%dispersion potential associated to the $\rho$, $\phi$ and $z$ polarizations can be
%written as
%
\begin{eqnarray}
\label{Vz_rr}
V_{a,z}^{rr}\left(\rho,\phi,q\right)\!\!\!\!&=&\!\!\!\! {1\over
4}\sum_b\hbar\Gamma_{a\rightarrow b}^{\,z}\Bigg\lbrace
 \sum_{l=0}^{q-1}U_{\parallel}^{rr}\Bigl(2\vert k_{ba}\vert\rho\sin\left(\phi +
 \pi l/q\right)\Bigr) \;\; -
 \cr
&{\;}&\;\;\;\;\;\;\;\;\;\;\;\; \;\;\;\;\;\;\;\;\;\;\;\;
 \;\;\;\;\;\;\;\;\;\;\;\;\;\; - \;\;\;\;
 \sum_{l=1}^{q-1}U_{\parallel}^{rr}\Bigl(2\vert
k_{ba}\vert\rho\sin\left(\pi l/q\right)\Bigr)\Bigg\rbrace,\;\;\;\;\;
\cr
%%%%%%%%%%%%%%%%%%%%%%%%%%%%%%%%%%%%%%%%%%%%%%%%%%%%%%%%%%%%%%%%%%%%%%%%%%
%
\label{Vfi_rr}
V_{a,\phi}^{rr}\left(\rho,\phi,q\right)\!\!\!\!&=&\!\!\!\!\! {1\over
4}\!\sum_b\hbar\Gamma_{a\rightarrow b}^{\,\phi}
 \Bigg\lbrace\!\sum_{l=0}^{q-1}W_{rr}^{\,\phi}\Bigl(2\vert k_{ba}\vert\rho,\phi +
 \pi l/q\Bigr) +
\!\!\sum_{l=1}^{q-1}W_{rr}^{\,\phi}\Bigl(2\vert k_{ba}\vert\rho,\pi
l/q\Bigr)\!\!\Bigg\rbrace\,,
\cr
%
%\cr
%%%%%%%%%%%%%%%%%%%%%%%%%%%%%%%%%%%%%%%%%%%%%%%%%%%%%%%%%%%%%%%%%%%%%%%%%%
%
\label{Vrho_rr}
V_{a,\rho}^{rr}\left(\rho,\phi,q\right)\!\!\!\!&=&\!\!\!\!\! {1\over
4}\!\sum_b\hbar\Gamma_{a\rightarrow b}^{\,\rho}
 \Bigg\lbrace\!\sum_{l=0}^{q-1}W_{rr}^{\,\rho}\Bigl(2\vert k_{ba}\vert\rho,\phi+\pi l/q\Bigr) -
\!\!\sum_{l=1}^{q-1}W_{rr}^{\,\rho}
 \Bigl(2\vert k_{ba}\vert\rho,\pi
 l/q\Bigr)\!\!\Bigg\rbrace\,,\nonumber
\end{eqnarray}
where
%we  defined the quantity
%
$
\Gamma_{a\rightarrow b}^{\,\sigma} =
\mbox{\large$\frac{4}{\hbar}$}\vert\langle a\vert d_{\sigma}\vert
b\rangle\vert^2\vert k_{ba}\vert^3
$
and the functions
%
%appearing
%
 in last equations are defined as
%
%\vskip -0.4 cm
%
\begin{eqnarray}
\label{Urrper}
U_{\perp}^{rr}(x)&=&\!\!\! -{1\over 2 x^3}\left(\cos x+x\sin
x\right)\; ;\;\;\;
%
%\cr
%%%%%%%%%%%%%%%%%%%%%%%%%%%%%%%%%%%%%%%%%%%%%%%%%%%%%%%%%%%%%%%%
\label{Urrpar}
U_{\parallel}^{rr}(x) = -{1\over 2 x^3}\left(\cos x+x\sin x-x^2\cos
x\right)\,,
\cr\cr
%%%%%%%%%%%%%%%%%%%%%%%%%%%%%%%%%%%%%%%%%%%%%%%%%%%%%%%%%%
%
\label{W1rr}
W_{rr}^{\,\phi}\left(x,\psi\right)\!\!\! &=&\!
U_{\parallel}^{rr}\left(x\sin\psi\right)
 \sin^2\psi+2U_{\perp}^{rr}\left(x\sin\psi\right)\cos^2\psi\,,
\cr\cr
%%%%%%%%%%%%%%%%%%%%%%%%%%%%%%%%%%%%%%%%%%%%%%%%%%%%%%%%%%
%
\label{W2rr}
W_{rr}^{\,\rho}\left(x,\psi\right)\!\!\! &=&\!
U_{\parallel}^{rr}\left(x\sin\psi\right)
 \cos^2\psi+2U_{\perp}^{rr}\left(x\sin\psi\right)\sin^2\psi\,.\nonumber
%
%%%%%%%%%%%%%%%%%%%%%%%%%%%%%%%%%%%%%%%%%%%%%%%%%%%%%%%%%%%%%%%%%
%
\end{eqnarray}
%
%

%\vskip -0.2 cm
 The $(fr)$ contribution to the level shift,
  $\delta E_a^{fr}$, can be obtained by a completely
  analogous procedure. This contribution depends on the
  state of the field. Here, we consider the field in the vacuum
state, $\langle n_{{\bf k}\lambda}\rangle = 0$.
In this particular case, it can be shown that
the $(fr)$ dispersion potential,
$V_{a,\sigma}^{fr}\left(\rho,\phi\right)$, may be obtained from
previous equations by the shortcut:
%
%\vskip -0.4cm
$$
U_{\parallel(\perp)}^{rr}(x)\longrightarrow
U_{\parallel(\perp)}^{fr}(x) =
U_{\parallel(\perp)}(x)-U_{\parallel(\perp)}^{rr}(x)\,,\;\;\;\;\;
 \sum_b\longrightarrow\sum_{b>a}-\sum_{b<a}\; ,
$$
%
%\vskip -0.3cm
 \noindent
 where functions $U_{\perp}$ and
$U_{\parallel}$ are defined by ($\gamma$ is the Euler-Mascheronni
constant)
%
%\vskip -0.6 cm
\begin{eqnarray}
\label{Uper}
U_{\perp}(x) &=& \!\!\!{1\over \pi
x^3}\left[-\mF\left(x\right)+x\mG\left(x\right)\right]\; ;\;\;\;
%
%\cr\cr
%%%%%%%%%%%%%%%%%%%%%%%%%%%%%%%%%%%%%%%%%%%%%%%%%%%%%%%%%%
%
\label{Upar}
U_{\parallel}(x) = {1\over \pi
x^3}\left[\left(x^2-1\right)\mF\left(x\right)+x\mG\left(x\right)-x\right]\,,
\cr
%%%%%%%%%%%%%%%%%%%%%%%%%%%%%%%%%%%%%%%%%%%%%%%%%%%%%%%%%%%%%%%%%%%%%%%%%%%%%%%%%
%
%%%%%%%%%%%%%%%%%%%%%%%%%%%%%%%%%%%%%%%%%%%%%%%%%%%%%%%%%%%%%%%%%
%
\label{F}
\mathcal F\left(x\right) &=& \int_0^{\infty}\dev t\;{\sin t\over
t+x}\;=\;\mathrm{C}i\left(x\right)\sin
x-\mathrm{s}i\left(x\right)\cos x\,,
\cr
%%%%%%%%%%%%%%%%%%%%%%%%%%%%%%%%%%%%%%%%%%%%%%%%%%%%%%%%%%%%%%%%%
%
\label{G}
\mathcal G\left(x\right) &=& -\int_0^{\infty}\dev t\;{\cos t\over
t+x}\;=\;\mathrm{C}i\left(x\right)\cos
x+\mathrm{s}i\left(x\right)\sin x\;=\;{\dev\over\dev x}\mathcal
F\left(x\right)\,,
\cr
%%%%%%%%%%%%%%%%%%%%%%%%%%%%%%%%%%%%%
%
\mathrm{s}i\left(x\right) &=& -{\pi\over 2}+\int_0^x \dev t\;{\sin
t\over t}\,,\;\;\;\;
\mathrm{C}i\left(x\right) = \gamma + \ln x + \int_0^x \dev t\;{\cos
t-1\over t}\,.\nonumber
\end{eqnarray}
The full dispersive potential of an atom in a state
 $\vert a\rangle$ is given by the sum of both
 $(rr)$ and  $(fr)$ contributions, so that
$
\label{fullV0}
V_a\left(\rho,\phi,q\right) =
\sum\!\!_{{\;}\atop{\sigma}}\left(V_{a,\sigma}^{rr}\left(\rho,\phi,q\right)
+
 V_{a,\sigma}^{fr}\left(\rho,\phi,q\right)\right)\,.
$
Let us, then, compute the van der Waals potential between the atom
and the wedge. For this case, the atom is in its ground state
$(a\rightarrow g)$, and assuming spherical symmetry,
 $ \vert\langle g\vert d_{z}\vert e\rangle\vert^2 =
\vert\langle g\vert d_{\phi}\vert e\rangle\vert^2 = \vert\langle
g\vert d_{\rho}\vert e\rangle\vert^2 = {1\over 3}\vert\langle g\vert
\mathbf{d}\vert e\rangle\vert^2\,,
 $
where the index $e$ specifies an excited state, we finally obtain
%
%\vskip -0.6cm
\begin{eqnarray}
\label{Vg}
V_{g}\left(\rho,\phi,q\right)\;\, =\;\, {1\over
2}\sum_e\hbar\Gamma_{e\rightarrow g}^{\mathrm{spt}}\Bigg\lbrace
 \sum_{l=0}^{q-1}
 \!\!\!\!\!\!&U&\!\!\!\!\!\!
  \Bigl(2 k_{eg}\rho \sin\left(\phi+\pi l/q\right)\Bigr)
  \;\;- \cr
  \sum_{l=1}^{q-1}\!\Big[U_{\parallel}\!\Bigl(2k_{eg}\rho\sin\left(\pi
l/q\right)\Bigr)\cos^2\left(\pi l/q\right)\;
 \!\!\!\!\!&-&\!\!\!\!
 U_{\perp}\!\Bigl(2k_{eg}\rho\sin\left(\pi
l/q\right)\Bigr)\cos\left(2\pi l/q\right)\!\Big]\!\Bigg\rbrace\;
\end{eqnarray}
where $U(x)=U_{\parallel}(x) + U_{\perp}(x)$ and
$ \Gamma_{e\rightarrow g}^{\,\mathrm{spt}} =
\mbox{\large$\frac{4}{3\hbar}$}\vert\langle g\vert\mathbf{d}\vert
e\rangle\vert^2 k_{eg}^3.
 $
This is the main result of the present paper. As a check of our
result, let us reobtain from the above equation a few known results
existent in the literature.
 Let us start with the atom-wall interaction. This
case corresponds to take $\phi_0=\pi$ ($q=1$). For any distance
regime, we get
\begin{equation}\label{Vg}
V_{g}\left(\rho,\phi,q=1\right) =  {1\over
2}\sum_e\hbar\Gamma_{e\rightarrow g}^{\mathrm{spt}}U\left(2
k_{eg}z\right)\;\;\;\; (z:=\rho\sin\phi)\, ,
\end{equation}
which was shown in \cite{TarciroJPA} to yield the well known results
in the literature. Further, in order to particularize the above
result for the non-retarded and retarded regimes we must take the
appropriate approximations of functions $ U_{\perp}$ and
$U_{\parallel}$. For the former (short distances) we have
$ U_{\perp}(x)\simeq U_{\parallel}(x) \simeq -
\mbox{\large$\frac{1}{2x^3}$}$
while for the latter (large distances), we have
$ U_{\parallel}(x)\simeq 2\, U_{\perp}(x) \simeq -
\mbox{\large$\frac{4}{\pi x^4}$} $.
Substituting into (\ref{Vg}) the short-distance behaviour of these
functions as well as relation $\Gamma_{e\rightarrow
g}^{\,\mathrm{spt}} = \mbox{\large$\frac{4}{3\hbar}$}\vert\langle
g\vert\mathbf{d}\vert e\rangle\vert^2 k_{eg}^3$, the non-retarded
atom-wall interaction is given by
\begin{equation}\label{Atom-Wall-NR-A}
V_{g}^{NR}(z) = -\frac{1}{12z^3} \sum_e \vert\langle g\vert{\bf
d}\vert
 e\rangle\vert^2
 \, .
 \end{equation}
Since we assumed spherical symmetry, we may write $\alpha_{ge}^\rho
= \alpha_{ge}^\phi=\alpha_{ge}^z=:\alpha_{ge}$, and from the
definition of $\alpha_{ab}^j$ given after equation (\ref{a'+-}), we
have $\alpha_{ge} = \mbox{\large$\frac{2}{3}$}\vert\langle
g\vert{\bf d}\vert e\rangle\vert^2/(\hbar c k_{eg})$ (recall that
$k_{ge} = -k_{eg}$). Hence, equation (\ref{Atom-Wall-NR-A}) can be
written in terms of $\alpha_{ge}$ as
\begin{equation}\label{Atom-Wall-NR-B}
V_{g}^{NR}(z) = -\frac{\hbar c}{8z^3} \sum_e \alpha_{ge}k_{eg}
 \, .
 \end{equation}
Substituting now in (\ref{Vg}) the large-distance behaviour of
functions $ U_{\perp}$ and $U_{\parallel}$, we get
 \begin{equation}\label{Atom-Wall-R}
V_{g}^{R}(z) = -\frac{3\hbar c}{8\pi z^4}
 \sum_e \alpha_{ge} =
-\frac{3\hbar c\,\alpha(0)}{8\pi z^4}\, ,
\end{equation}
where $\alpha(0)$ is the static polarizability of the atom.

For  an atom in the region between two parallel conducting plates,
discussed in detail by Barton \cite{Barton1987}, we were not able to
establish an equivalence analytically. However, we have checked
numerically that both results are in complete agreement.

Now, as our last particular case, we reobtain the retarded potential
for the atom-wedge system calculated by Brevik, Lygren and
Maracheviski \cite{BrevikEtAl1998}. For large distances, our general
result (\ref{Vg}) takes the form (this is called the Casimir-Polder
regime)
$$
V_{CP}(\rho,\phi,q) = {\hbar\over
32\pi\rho^4}\sum_e{\Gamma_{e\rightarrow g}^{\mathrm{spt}}\over\vert
k_{eg}\vert^4}\left[-6\sum_{l=0}^{q-1}\sin^{-4}\left(\phi+\pi
l/q\right)+2\sum_{l=1}^{q-1}\sin^{-4}\left(\pi l/q\right)\right]\;.
$$
Previous summations can be evaluated and the previous result can be
cast into the form
$$
V_{CP}(\rho,\phi,q) = {\hbar c\,\alpha(0)\over
4\pi\rho^4}\left[{1\over
90}\left(q^2-1\right)\left(q^2+11\right)-{q^2\over
\sin^2q\phi}\left({3q^2\over 2\sin^2q\phi}+1-q^2\right)\right]\, ,
$$
considered now valid for every value of $q$, in perfect agreement
with \cite{BrevikEtAl1998}.

\vspace{-20pt}

\begin{figure}[!h]
\begin{center}
\includegraphics[width=3.4in]{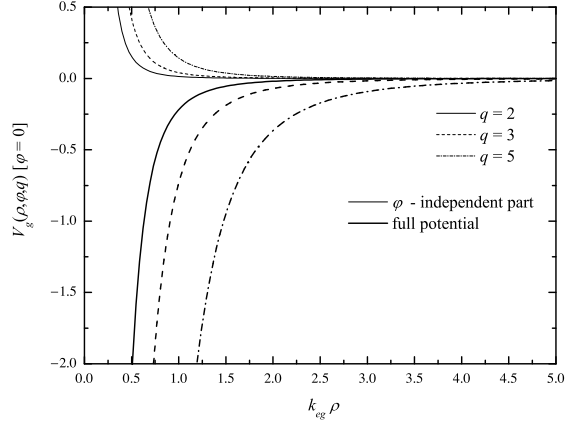}
\label{FigVrho}
 \vspace{-20pt}
 \caption{$V_g(\rho,\varphi=0,q)$ {\it
versus} $\rho$ for different values of $q$. For increasing $q$
(closing the wedge) the interaction become stronger, as expected.}
\end{center}
\end{figure}

 Figure \ref{FigVrho} shows the variation of $V(\rho,\phi,q)$,
   given by equation (\ref{Vg}), in terms of $\rho$ for fixed $\phi$,
   taken conveniently as   $\phi = \phi_0/2$ ($\varphi=0$) and for
 different values of $q$. The full potential is represented by the strong
 lines: the solid one corresponding to $\phi_0=\pi/2$, the
 dashed one to $\phi_0=\pi/3$ and the dotted-dashed one to
 $\phi_0=\pi/5$. The component of the force along $\hat\rho$, denoted by $F_\rho$, is
 attractive. However, looking at equation (\ref{Vg}), we note that
 the last two contributions are independent of $\phi$ and with a repulsive
 character. In other words, no matter the value of $\phi$, there is always
 a repulsive corner contribution to the  force $F_\rho$, represented
 in Figure \ref{FigVrho} by the thin lines.

\begin{figure}[!h]
\begin{center}
\includegraphics[width=3.4in]{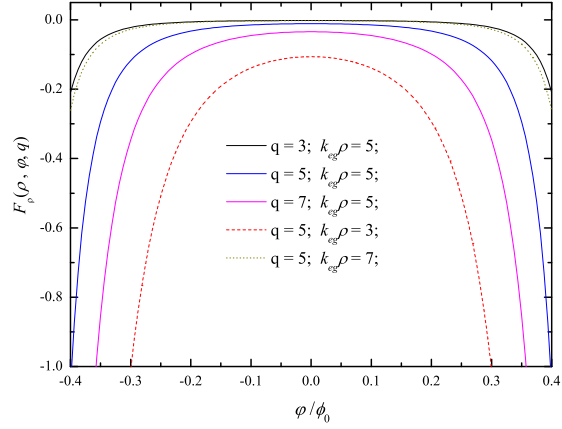}
\label{A}
 \vspace{-20pt}
  \caption{$F_\rho(\rho,\varphi,q)$ with
fixed $\rho$ {\it versus} $\varphi/\phi_0$ for different values of
$q$ and $\rho$.}
%($\ne$ angles of the wedge).}
\end{center}
\end{figure}

 Figure 3  shows $F_\rho = -
  \frac{\partial V_g}{\partial\rho}$ in terms of $\phi$ for fixed $\rho$
  and for  different values of $q$. Note that the modulus of $F_\rho$ increases
  as we get closer and closer to each plate. Even for $\phi=\phi_0/2$ ($\varphi=0$), i.e.,
  for  points equidistant of both plates forming the wedge, $F_\rho$ is not zero, though
  it assumes  minimum values at these points. Observe, also, that
  for $\varphi=0$ the modulus of $F_\rho$ increases as the angle
  between the plates diminishes. Naively, one could think this is a
  paradoxal result, since the plates are becoming more and more
  parallel to each other and for an atom inside two parallel plates
  there is no component of the force parallel to the plates.
  However, to obtain correctly the limit of an atom inside two
  parallel plates one must not only diminish the angle between the
  plates but also take $\rho$ to infinite, with the constraint
  $\rho\phi_0=a$, $a$ being the distance between the parallel
  plates. A simple but careful analysis shows that, in fact,
  $F_\rho(\rho,\varphi=0,q)$ increases if we mantain $\rho$ fixed
  and finite and increase $q$ (close the wedge).

\vspace{-20pt}

\begin{figure}[!h]
\begin{center}
\includegraphics[width=3.4in]{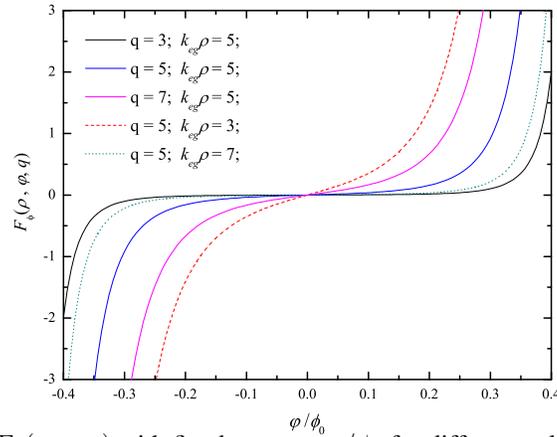}
\label{B}
 \vspace{-30pt}
 \caption{$F_\phi(\rho,\varphi,q)$ with fixed $\rho$ {\it versus} $\varphi/\phi_0$
for different values of $q$ and $\rho$.}
%($\ne$ angles of the wedge).}
\end{center}
\end{figure}

Figure 4 shows $F_\phi = - \mbox{\large$\frac{1}{\rho}\frac{\partial
V_g}{\partial\phi}$}$ in terms of $\phi$ for fixed $\rho$
  and for  different values of $q$. In contrast to what happens to $F_\rho$, $F_\phi=0$
 for $\varphi=0$. Besides,
  $F_\phi<0$ for negative $\varphi$ and   $F_\phi>0$ for positive
  $\varphi$: except for $\varphi=0$, the atom is attracted to the
 nearest plate (apart from a $F_\rho$ component
 pointing to the corner)

\section{Final comments and perspectives}

Using an approach based on a master equation, we computed the
dispersion force on an electrically polarizable atom near a
perfectly conducting wedge. Our result is valid for both retarded
and non-retarded regimes, generalizing in this way a previous result
of the literature \cite{BrevikEtAl1998}, which is valid only in the
retarded regime. We  checked  our calculations by reobtaining well
known results in the literature, as for instance, the non-retarded
and retarded interactions between an atom and a perfectly conducting
wall, given by equations (\ref{Atom-Wall-NR-A}) (or equivalently
(\ref{Atom-Wall-NR-B})) and (\ref{Atom-Wall-R}). It is worth noting
that since the non-relativistic expression $\vert\langle g\vert{\bf
d}\vert e\rangle\vert^2$ does not depend on $c$, the non-retarded
potential  is independent of c (see equation
(\ref{Atom-Wall-NR-A})), as it should be (the dependence on $\hbar$
comes from $\vert\langle g\vert{\bf d}\vert e\rangle\vert^2$). In
 (\ref{Atom-Wall-NR-B}), the dependence in $c$ is just apparent,
 since it cancels with that appearing in $k_{eg}=\omega_{eg}/c$.
 We should emphasize, however, that in the non-retarded regime
  expression (\ref{Vg}) is valid only for $\phi_0=\pi/q$, with
$q$ a positive integer number and $\phi_0$ being the angle of the
wedge. We found a full $\rho$-component of the force which is
attractive. There is, however, a $\phi$-independent repulsive
contribution for $F_\rho$, which we interpreted as a corner
contribution. A numerical analysis should be made to see if our
results can be of some relevance in future experiments using this
geometry. The method employed here can be applied in the computation
of resonant potentials (excited atoms), but this will be left for a
future work. Our formalism can still be used in the computation of
spontaneous emission rates of an atom near boundaries as the one
considered here.

\noindent {\bf Acknowledgments:}FSSR and CF are very indebted to G.
Barton, R. Passante, C. Villarreal and R. Rodrigues for valuable
 comments. FSSR  thanks M. Moriconi for the kind hospitality
at UFF, where part of this work was done, and FAPERJ for financial
support. TCNM and CF acknowledges FAPERJ and CNPq for financial
support.

%
%
%%%%%%%%%%%%%%%%%%%%%%%%%%%%%%%%%%%%%%%%%%%%%%%%%%%%%%%%%%%%%%%%%%%%%%%%%%%%%%%%%%%
%
%

\end{document}